\documentstyle[12pt]{article}

\newcommand\mathC{{\mkern1mu\raise2.2pt\hbox{$\scriptscriptstyle|$}
        {\mkern-7mu\rm C}}}
\newcommand{\mathR}{{\rm I\! R}}                

\renewcommand\[{[\,}                            
\renewcommand\]{\,]}                            

\renewcommand\mathR{{\rm I\! R}}

\newcommand\unit{{\rm 1\kern-3.2pt I}}

\begin{document}
\title{General relativity histories theory: \\
Spacetime diffeomorphisms and the Dirac algebra of constraints}
\author{ Ntina Savvidou \thanks{ntina@ic.ac.uk}\\ {\small Theoretical
Physics Group, The Blackett Laboratory,} \\ {\small Imperial College,
 SW7 2BZ, London, UK} \\ }

\maketitle

\begin{abstract}

We show that representations of the group of spacetime diffeomorphisms {\em and\/} the
Dirac algebra {\em both\/} arise in a phase-space histories version of canonical
general relativity. This is the general-relativistic analogue of the novel time
structure introduced previously in history theory: namely, the existence in
non-relativistic physics of two types of time translation; and the existence in
relativistic field theory of two distinct Poincar\'{e} groups.
\end{abstract}

\renewcommand {\thesection}{\arabic{section}}
 \renewcommand {\theequation}{\thesection.\arabic{equation}}
\let \ssection = \section
\renewcommand{\section}{\setcounter{equation}{0} \ssection}

\pagebreak

\section{Introduction}
The `problem of time' in quantum gravity takes on a different form according to the
approach to quantum gravity that one adopts. However, in all cases, this question of
the status of the notion of time is a fundamental one; indeed, it is true in all cases
that we need a better understanding of conceptual issues concerning the nature of
space and time.

In light of the recent developments on the distinction that has been made between time
as a causal ordering parameter, and time as an evolution parameter in dynamics
\cite{Sav99}, the main goal of the present paper is to show how the application of
history ideas  to general relativity opens up a novel way of viewing that subject; and
hence ultimately to a completely new way of tackling the quantisation of gravity.

In what follows, we apply the ideas of classical history theory \cite{Sav99} to the
general theory of relativity. A preliminary step in this direction was the application
of the history methods to parametrised systems (often used as simple models for
general relativity). Thus, in \cite{SA00} we studied the quantisation of constrained
systems using the continuous-time histories scheme. In particular, the existence of
the two times in a history version of parametrised systems was exploited to show the
existence of an intrinsic time that does not disappear when the constraints are
enforced, either classically or quantum mechanically. Hence this provides a solution
to the `problem of time' for systems of this type. This work is a natural precursor
for dealing with the problem of time as it appears in canonical quantum gravity.

In the context of general relativity, we start by considering a Lorentzian geometry on
a spacetime $M\simeq \mathR\times\Sigma$ as being equivalent to a history of
Riemannian metrics on the three-manifold $\Sigma$. Thus we consider paths $t\mapsto
h_{ij}(t,\underline x)$ of Riemannian metrics, which together with the paths $t\mapsto
\pi^{kl}(t,\underline x)$ of conjugate momenta, are postulated to form the fundamental
classical history algebra (the history analogue of the normal canonical Poisson
brackets)
\begin{eqnarray}
\{ h_{ij}(t , \underline{x})\,, h_{kl} ( t^{\prime} , \underline{x}^{\prime} ) \} &=&
0 \label{GR1}\\
\{ \pi^{ij} (t , \underline{x})\,, \pi^{kl} ( t^{\prime} , \underline{x}^{\prime} ) \}
&=& 0 \label{GR2}\\
\{ h_{ij} (t , \underline{x})\,, \pi^{kl} ( t^{\prime} , \underline{x}^{\prime} ) \}
&=& {\delta}_{(ij)}^{kl}\, \alpha(t^{\prime}) \delta ( t , t^{\prime})\, \delta^3 (
\underline{x} , \underline{x}^{\prime}), \label{GR3}
\end{eqnarray}
where we define ${{\delta}_{(ij)}}^{kl} := \frac{1}{2}({\delta_i}^k {\delta_j}^l +
{\delta_i}^l {\delta_j}^k )$, and where $\alpha(t)$ is some strictly positive scalar
density of weight -1 in the variable $t$.

In the standard approach to canonical general relativity, the relation between the
spacetime diffeomorphisms algebra and the Dirac constraint algebra has long been an
important matter for discussion \cite{Kuc91,I92}. Therefore, it is of considerable
significance that in this new construction the two algebras appear together for the
first time in a completely natural way: specifically, as we shall show, the history
theory contains a representation of both the spacetime diffeomorphisms group {\em
and\/} the Dirac algebra of constraints of the canonical theory.

In particular, we augment the history space of the canonical general relativity
treatment, to define appropriate covariant Poisson brackets. A key result in this
respect is the observation that for each vector field $W$ on $M$, the `Liouville'
function $V_W$ that is defined as
\begin{equation}
    V_W:=\int d^4X\,\pi^{\mu\nu}(X){\cal L}_W g_{\mu\nu}(X)
\end{equation}
satisfies the Lie algebra of the group of spacetime diffeomorphisms
\begin{equation}
\{\, V_{W_1}, V_{W_2}\,\} = V_{ [\, W_1 , W_2\,]},
\end{equation}
for all spacetime vector fields $W_1, W_2$, and where $[\, W_1 , W_2\,]$ denotes their
commutator.

This is a very significant result since it implies that in this history theory there
is a central role for {\em spacetime\/} concepts, whereas the canonical approaches to
general relativity are dominated by spatial ideas. Furthermore, it makes particularly
clear how the distinction between the Dirac constraint algebra and the spacetime
diffeomorphisms group arises as a facet of the non-trivial temporal structure of the
histories description.

In section 2 we present the basic ideas of the histories temporal structure. Of
particular importance is the distinction between the aspects of the concept of time as
(i) an ordering parameter, and (ii)  an evolution parameter. This is  realised
mathematically with the construction of two distinct generators of time translations.
Furthermore, the definition of the {\em action operator S}---which was proved to be
the generator of both types of time translations---nicely intertwines the two modes of
time \cite{Sav99}. We briefly present the histories classical non-relativistic physics
and relativistic field theory. We especially emphasise the existence of two
Poincar\'{e} groups as the analogue of the two types of time translation in
non-relativistic physics.\cite{Sav01}. The histories theory of parameterised systems
is a natural precursor for the study of histories general relativity theory. We recall
how the existence of the two modes of time for such systems has as a result an
`intrinsic time' that does not disappear when enforcing the constraints \cite{SA00}.
This provides a solution to the analogue for the parametrised particle of the famous
`problem of time' in canonical quantum gravity.

In section 3 we first present the structure of the history version of canonical
general relativity theory. We explicitly write the analogue of the Dirac algebra of
constraints. We then show that---after appropriately augmenting the history space of
canonical histories---there exists a representation of the group of spacetime
diffeomorphisms. This novel result is a direct analogue of the two Poincar\'{e} groups
in relativistic field theory: it is therefore grounded in the distinction between the
two aspects of time. Next we explicitly write the extended canonical history algebra
and we show that the original history state space is the state space of the standard
$3+1$ decomposition.

Finally we apply the history ideas for the treatment of parameterised systems and we
show that, of all spacetime diffeomorphisms it is only the generators of time
reparameterisations---{\em i.e.\/} the ones related to the Liouville function
$V$---that are still defined on the histories reduced state space.

\vspace{1.2cm}

\section{Background}

\subsection{Temporal Structure of HPO histories theory}
In recent years, it has become better understood that the problem of quantum
gravity---and especially the way in which time might appear in such a
theory---suggests the need for a new form of quantum theory: in particular, one where
the notion of `time' is introduced in some novel way.

One such formalism is the `HPO' (`History Projection Operator') approach to a quantum
history theory. Although the programme originated from the consistent histories
theory, as formulated initially by Griffiths, Omn\`{e}s, and Gell-Mann and Hartle
\cite{CoHis}, it was developed so that the logical structure of the histories theory
was recovered\footnote{A crucial problem of the consistent histories theory is that
the original definition of a history has as a consequence the loss of the single-time
quantum logical structure }; in particular it introduced a `temporal' logic of the
theory \cite{cji94}. The HPO theory takes a completely different turn in the way the
concept of time was introduced in \cite{Sav99}.

The consistent histories formalism  was developed to deal with closed systems. A
history $ \alpha = ( \hat{\alpha}_{t_{1}},\hat{\alpha}_{t_{2}},\ldots,
\hat{\alpha}_{t_{n}})$ is defined to be a collection of projection operators
$\hat\alpha_{t_i}$, $i=1,2,\ldots,n$, each of which represents a property of the
system at the single time $t_i$. Therefore, the emphasis is placed on histories,
rather than properties at a single time, which in turn gives rise to the possibility
of generalized histories with novel concepts of time.

The History Projection Operator approach, developed originally by Isham \cite{cji94},
and Isham and Linden \cite{HPOHis}, is an approach to the consistent histories
formalism that places emphasis on temporal logic. This is achieved by representing the
history $\alpha$ as the operator $\hat\alpha :=\hat{\alpha}_{t_{1}}\otimes\hat{\alpha}
_{t_{2}}\otimes\cdots \otimes \hat{\alpha}_{t_{n}}$ which is a genuine {\em
projection\/} operator on the tensor product $ \otimes_{i=1}^n {\mathcal{H}}_{t_i} $
of copies of the standard Hilbert space $\mathcal{H}$. Note that to use this
construction in any type of field theory requires an extension to a continuous time
label, and hence to an appropriate definition of the continuous tensor product
$\otimes_{t\in\mathR}{\mathcal{H}}_t$. This has been done successfully for
non-relativistic particle physics \cite{HPOHis}, and relativistic quantum field theory
\cite{Sav01}.

A central feature of the HPO histories theory is the development of the novel temporal
structure that was introduced in \cite{Sav99}. Specifically, it was shown that there
exist {\em two\/} distinct types of time transformation, each of which represents a
distinct quality, or mode, of the concept of time.

The first such mode corresponds to time considered purely as a kinematical parameter
of a physical system, with respect to which a history is defined as a succession of
possible events. It is strongly connected with the temporal-logical structure of the
theory and it is related to the view of time as a parameter that determines the
ordering of events. The second mode corresponds to the dynamical evolution generated
by the Hamiltonian.

Classically, these two ways of considering time are nicely intertwined through the
histories analogue of the action principle which provides the paths that are solutions
to the classical equations of motion. A main result of the theory is that physical
quantities appear naturally time-averaged. Hence these new ideas on the concept of
time have as a consequence that observables admit {\em two\/} different time labels:
(i) a time parameter $t$ which corresponds to the `external' time that labels events
at different moments of time, and with respect to which the time averages are taken;
and (ii) a time parameter $s$ which corresponds to the `internal' time that appears as
the evolution parameter for a fixed external time $t$.

In the corresponding quantum theory, the Hamiltonian\footnote{In this context,
`Hamiltonian' $H = \int\!dt \, H_t $ means the history quantity that is the
time-averaged energy of the system.} operator $H$, and the `Liouville' operator $V$
are the generators of the two types of time transformation \cite{Sav99}. Specifically,
the Hamiltonian $H$ is the generator of the unitary time evolution with respect to the
`internal' time label $s$; this has no effect on the `external' time label $t$. On the
other hand, the Liouville operator $V$---defined in analogy to the kinematical part of
the classical action functional---generates time translations along the $t$-time axis
without affecting the $s$-label.

The key feature of the ensuing temporal structure, however, is the definition of the
\textit{action} operator $S$ as a quantum analogue of the classical action functional:
\begin{equation}
S:= V - \int^{+\infty}_{-\infty}\!dt\,H_{t}\, = \,V -H.
\end{equation}
It transpires that the action operator $S$ generates \textit{both} types of time
transformation, and in this sense it is the generator of {\em physical} time
translations in the HPO formalism.

The time transformations generated by the action operator $S$ resemble the canonical
transformations generated by the Hamilton-Jacobi action functional. In this sense,
there is an interesting relation between the definition of $S$ and the well-known work
by Dirac on the Lagrangian theory for quantum mechanics \cite{Dirac,Sav99}. In
particular, motivated by the fact that---contrary to the Hamiltonian method---the
Lagrangian method can be expressed relativistically (on account of the action function
being a relativistic invariant), Dirac tried to take over the general $\it{ideas}$ of
the classical Lagrangian theory, albeit not the equations of the Lagrangian theory
{\em per se\/}.

Recently, these ideas have been applied in various theories, with some intriguing
results. For example, the temporal structure of HPO histories enables us to treat
parameterised systems in such a way that the problem of time does not arise
\cite{SA00}. Indeed, histories keep their intrinsic temporality after the
implementation of the constraint: thus there is no uncertainty about the
temporal-ordering properties of the physical system.

In relativistic quantum field theory,  the analogue of the two types of time
transformation is the existence of two  groups of \textit{Poincar\'{e}}
transformations \cite{Sav01}. It transpires that different representations of the
theory---that correspond to different choices of foliation---can all be defined on the
{\em same\/} Hilbert space, and they are related by transformations generated by the
`external' Poincar\'{e} group.

As we shall see in what follows, the histories description of general relativity
blends together the structure of the two systems referred above: namely, the
parameterised systems and the relativistic field.

\subsection{Classical Histories}
In the histories formalism for Newtonian classical mechanics, the space of classical
histories $ \Pi = \{ \gamma \mid \gamma : \mathR \rightarrow \Gamma \}$ is the set of
all smooth paths on the classical state space $\Gamma$. It can be equipped with a
natural symplectic structure, which gives rise to the Poisson bracket
\begin{eqnarray}
    \{x_t\, , x_{t^{\prime}} \}_{\Pi} &=& 0\\
    \{p_t\, , p_{t^{\prime}} \}_{\Pi} &=& 0\\
  \{x_t\, , p_{t^{\prime}} \}_{\Pi} &=& \delta (t-t^{\prime})
  \end{eqnarray}
where
\begin{eqnarray}
x_t : \Pi & \rightarrow & \mathR  \\
\gamma & \mapsto & x_t (\gamma) := x( \gamma (t))
\end{eqnarray}
and similarly for $ p_t $.

The classical analogue of the Liouville operator is defined as
 \begin{equation}
  V(\gamma) := \int_{-\infty}^\infty \! dt\, p_t\, \dot{x}_t ,
 \end{equation}
and the Hamiltonian ({\em i.e.}, time-averaged energy) function $H$ is defined as
\begin{equation}
  H ( \gamma ):= \int_{-\infty}^{\infty}\!dt \,H_t(x_t,p_t)
\end{equation}
where $H_t$ is the Hamiltonian that is associated with the copy ${\Gamma}_t$ of the
normal classical state space with the same time label $t$.

The temporal structure leads to the histories analogue of the classical equations of
motion
\begin{equation}
\{ F , V \}_{\Pi}\, (\gamma_{cl}) = \{ F , H \}_{\Pi}\, (\gamma_{cl})
\end{equation}
where $F$ is any function on $\Pi$, and where the path $\gamma_{cl}$ is a solution of
the equations of motion.

A crucial result therefore is that, the history equivalent of the classical equations
of motion is given by the following condition that holds for \textit{all} functions
$F$ on $\Pi$ when $\gamma_{cl}$ is a classical solution:
\begin{equation}
\{F , S\}_{\Pi}\, (\gamma_{cl}) = 0,
\end{equation}
where
\begin{equation}
S( \gamma ) := \int_{-\infty}^{\infty}\!dt \, (p_t\dot{x_t} - H_t(x_t,p_t))=\, V(
\gamma )- H( \gamma )
\end{equation}
is the classical analogue of the action operator. This is the history analogue of the
least action principle \cite{Sav99}.

\subsection{Classical Parameterised Systems}
A natural precursor to general relativity is the theory of parameterised systems. Such
systems have a vanishing Hamiltonian $H = h(x , p) $, when the constraints are
imposed. Classically this implies that two points of the constraint surface $C$
correspond to the same physical state; hence the true degrees of freedom are
represented by points in the reduced state space ${\Gamma}_{red}$
\begin{equation}
{\Gamma}_{red} := C / \sim
\end{equation}
An element of the reduced state space is {\em itself \/} a solution to the classical
equations of motion; on the other hand, a point in state space also corresponds to a
possible configuration of the physical system at an instant of time. Hence the notion
of time is unclear: in particular, it is not obvious how to recover the notion of
temporal ordering unless we choose to arbitrarily impose a gauge-fixing condition.

In the histories approach to parameterised systems, the history constraint surface
$C_h$ is defined as $C_h = \{ {\gamma}_{c} : \mathR \rightarrow C \}$---the set of all
smooth paths from the real line to the constraint surface $C$. The history Hamiltonian
constraint is defined by $H_{\kappa} = \int \! dt\, \kappa (t) h_t$, where $h_t :=
h(x_t, p_t)$ is first-class constraint. For all values of the smearing function
$\kappa (t)$ the history Hamiltonian constraint $H_{\kappa}$ generates canonical
transformations on the history constraint surface $C_h$. The history reduced state
space $\Pi_{red}$ is then defined as $\Pi_{red} = \{\gamma: \mathR
\rightarrow\Gamma_{red}\}$---the set of all smooth paths on the canonical reduced
state space $\Gamma_{red}$: it is identical to the space of orbits of $H_{\kappa}$ on
$C_h$.

The novel result here is that, contrary to what is the case for existing treatments of
parameterised systems, the classical equations of motion \textit{can} be explicitly
realised on the reduced state space $\Pi_{red}$. They are given by
\begin{equation}
\{ \tilde S , F \}\,(\gamma_{cl}) = \{ \tilde{V} , F \}\,(\gamma_{cl}) = 0
\end{equation}
where $\tilde{S}$ and $\tilde{V}$ are respectively the action and Liouville functions
projected on $\Pi_{red}$.

Both $\tilde{S}$ and $\tilde{V}$ commute weakly with the Hamiltonian constraint
\cite{SA00}. Furthermore, the smeared form of the Liouville function $V_{\lambda}=
\int \!dt\, \lambda (t)\, p_t \,{\dot{x}}_t$ generates time reparameterisations on
$\Pi_{red}$, and it leaves invariant the classical equations of motion.

\subsection{Classical Field Theory}

We write the history version of classical field theory for Minkowski spacetime,
foliated with respect to a time-like vector $n^{\mu}$, that is normalised by
$\eta_{\mu\nu} n^{\mu} n^{\nu} = 1 $. We shall take the signature of the Minkowski
metric $\eta^{\mu\nu}$ to be $(+,-,-,-)$.

In the histories formalism of a scalar field, the space of state-space histories $\Pi$
is an appropriate subset of  the continuous Cartesian product $\times_t \Gamma_t $ of
copies of the standard state space $\Gamma$, each labeled by the time parameter $t$.
The choice of $\Gamma$ depends on the choice of a foliation vector $n^{\mu}$, hence
the space of histories also has an implicit dependence on $n^{\mu}$ and should
therefore be written as $^n \Pi$.

Furthermore, for each space-like surface $\Sigma_t = (n,t)$---defined with respect to
its normal vector $n$, and labeled by the parameter $t$---we consider the state space
$\Gamma_t = T^* C^\infty (\Sigma_t)$ that is defined in such a way as to give the
basic Poisson algebra relations of the history theory:
\begin{eqnarray}
\{\,\phi(X)\,, \phi(X^{\prime})\,\} &=& 0    \\
\{\, \pi(X)\,, \pi(X^{\prime}) \,\} &=& 0     \\
\{\, \phi(X)\,, \pi(X^{\prime}) \,\} &=& \delta^4 (X- X^{\prime})
\end{eqnarray}
where $X$ and $X^{\prime}$ are spacetime points. Note that a spacetime point $X$ can
be associated with the pair $(t,\underline{x})\in\mathR\times\mathR^3 $ as $X=tn +
x_n$, where the three-vector $\underline{x}$ has been associated with a corresponding
four-vector $x_n$ that is $n$-spatial ({\em i.e.}, $n\cdot x_n=0$); note that
$t=n\cdot X$.

We then define the action, Liouville and Hamiltonian functionals for the scalar field
as
\begin{eqnarray}
{}^n\!S&:=& {}^n\!V - {}^n\!H \\
{}^n\!V&:=& \int \!d^4 X\, \pi(X)\,n^{\mu} \partial_{\mu}\, \phi(X)  \\
{}^n\!H&:=& \frac{1}{2} \int \!d^4 X \, \left(\, \pi^2 (X) + \phi (X)\,{}^{n}\Gamma
\,\phi (X)\,\right) ,
\end{eqnarray}
respectively. Here ${}^n\Gamma := (n^{\mu} n^{\nu} - \eta^{\mu \nu})\partial_{\mu}
\partial_{\nu} +\tilde{m}^2$, where $\tilde{m}$ is the mass of the free field.

It can be shown that the variation of the action functional ${}^n\!S\[\gamma \]$
leaves invariant the paths $\gamma_{cl}$ that are the classical solutions of the
system:
\begin{eqnarray}
\{\phi(X), {}^n\!S \}(\gamma_{cl})&=& 0  \label{clft1} \\
\{\pi(X), {}^n\!S \}(\gamma_{cl})&=& 0. \label{clft2}
\end{eqnarray}

\subsubsection{Poincar\'{e} symmetry}
For each copy $\Gamma_t$ of the standard state space, there exists a Poincar\'{e}
group, as one would expect in a canonical treatment of relativistic field theory. On
the other hand, in histories theory the state space $\Pi$ is, heuristically, the
Cartesian product of such copies. Hence, for each copy of the standard state space,
labeled by a fixed value of $t$, there exists an `internal' Poincar\'{e} group acting
on the copy of standard canonical field theory, that is labeled with the same time
label $t$. However, the physical quantities in histories theory appear naturally
time-averaged \cite{Sav99}, and hence a central role is played by a time-averaged form
of these internal groups.

Of special interest is the action of the corresponding Hamiltonian
$^n{}\!H:=\int_{-\infty}^\infty dt\;{}^n\!H_t$, and the boost generator ${}^n\!K$, on
the field $\phi(X)=\phi(t, \underline{x})$. In particular, we can define a classical,
history analogue of the Heisenberg picture fields $\phi(X,s)=\phi(t , \underline{x},
s)$ \cite{Sav01} as
\begin{eqnarray}
\phi(X)
\begin{array}{c}
  {}^n\!H \\
  \longrightarrow\\
\end{array}
\phi(X,s)
\end{eqnarray}
or
\begin{eqnarray}
\phi(t,\underline{x})
\begin{array}{c}
  {}^n\!H \\
  \longrightarrow\\
\end{array}
\phi(t,\underline{x},s): = \cos({}^n\Gamma^{\frac{1}{2}}s)\, \phi(X) +
\frac{1}{{}^n\Gamma^{\frac12}} \sin( {}^n\Gamma^{\frac{1}{2}}s) \,\pi(X).
\end{eqnarray}

The action of boost transformations is best shown upon objects $\phi(X,s)$ as
\begin{equation}
\phi(t,\underline{x},s)
\begin{array}{c}
  {}^n\!K \\
  \longrightarrow\\
\end{array}
\phi(t, \underline{x}',s'),
\end{equation}
where $(\underline{x}', s')$ and $(\underline{x},s)$ are related by a Lorentz boost.

In addition to these `internal' Poincar\'{e} groups (and the time-averaged version)
there exists  an `external' Poincar\'{e} group with the same space translations and
rotations generators as those of the internal Poincar\'{e} group, but with different
time translator and boosts. In particular, the time-translation generator for the
`external' Poincar\'{e} group is the Liouville functional ${}^n\!V$ \cite{Sav01}:
\begin{eqnarray}
\phi(t,\underline{x})
\begin{array}{c}
  {}^n\!V \\
  \longrightarrow\\
\end{array}
\phi(t+\tau ,\underline{x}).
\end{eqnarray}

The boost generator ${}^n\!\tilde{K}(m)$ generates Lorentz transformations
\begin{eqnarray}
\phi(X) \rightarrow \phi(\Lambda X) \\
\pi(X) \rightarrow \pi(\Lambda X)
\end{eqnarray}
where $\Lambda$ is the element of the Lorentz group parameterised by the boost
parameter $m^i$.

Furthermore under the action of the external Poincar\'{e} group, the action functional
$^n\!S$ transforms as
\begin{equation}
^nS \;\;\;\rightarrow \;\;\;^{\Lambda n}\!S.
\end{equation}

It can be shown from Eqs.\ (\ref{clft1}--\ref{clft2}) that the two types of boost
transformation coincide for the classical solutions $\gamma_{cl}$ \cite{Sav01}
\begin{eqnarray}
\{ \phi(X), K(m) \} (\gamma_{cl}) &=&  \{ \phi(X), \tilde{K}(m)
\} (\gamma_{cl}) \\[4pt]
\{ \pi(X), K(m) \} (\gamma_{cl}) &=& \{ \pi(X), \tilde{K}(m) \} (\gamma_{cl}).
\end{eqnarray}

\vspace{1.2cm}

\section{Histories version of General Relativity}
In order to apply the histories theory to general relativity two methods may be
followed. From the spacetime perspective advocated by Hartle \cite{CoHis}, a history
is a Lorentzian metric. On the other hand, from the canonical perspective, a history
is a path in the space of Riemannian metrics on a fixed three-manifold $\Sigma$. We
shall start by following the latter approach here.

Another interesting way of formulating general relativity histories is a
covariant-like treatment---similar to the one developed by Wald \cite{Wald}---that
provides a clarifying spacetime description of the theory. This will be relevant in
future work, where we study the change of foliation in a covariant description of
histories theory.

\bigskip

\subsection{Canonical treatment: basic structure}
The history space $\Pi$ for general relativity is a suitable subset of the Cartesian
product ${\times}_t {\Gamma}_t $ of copies of the classical general relativity state
space $\Gamma = \Gamma(\Sigma)$, labeled by a parameter $t$, with $t\in \mathR$. Here
$\Sigma$ is a fixed three-manifold.

In particular, $\Gamma (\Sigma) = T^{*}{\rm Riem}(\Sigma)$, where ${\rm Riem}(\Sigma)$
is the space of Riemannian metrics on $\Sigma$;  {\em i.e.}, an element of $\Gamma
(\Sigma)$ is a pair $ ( h_{ij}, \pi^{kl})$. A history is defined to be any smooth map
$t\mapsto (h_{ij}(t,\underline{x}),\pi^{kl}(t,\underline{x}))$.

The history version of the canonical Poisson brackets is postulated---in accord to the
histories ideas\cite{Sav99,SA00,Sav01}, where the entries of the history algebra are
defined as histories, {\em i.e.,} paths of an appropriate history space---to be
\begin{eqnarray}
\{h_{ij} (t,\underline{x})\,, h_{kl} ( t^{\prime} ,
\underline{x}^{\prime} ) \} &=& 0 \label{GR1b}\\
\{ \pi^{ij} (t,\underline{x})\,, \pi^{kl} ( t^{\prime} ,
\underline{x}^{\prime} ) \} &=& 0 \label{GR2b} \\
\{ h_{ij} (t,\underline{x})\,, \pi^{kl} (t^{\prime} , \underline{x}^{\prime} )\} &=&
{{\delta}_{(ij)}}^{kl}\, \alpha(t')\delta ( t , t^{\prime})\, \delta^3 ( \underline{x}
, \underline{x}^{\prime}) \label{GR3b}
\end{eqnarray}
where we have defined ${{\delta}_{(ij)}}^{kl} := \frac{1}{2}({\delta_i}^k {\delta_j}^l
+ {\delta_i}^l {\delta_j}^k )$ and where $\alpha(t)$ is some strictly positive scalar
density of weight -1 in the variable $t$.

The appearance of  $\alpha(t)$ on the right-hand side of the canonical Poisson
brackets can be justified in the following way. The quantity
$\pi^{kl}(t,\underline{x})$ is a density in the spatial variable
$\underline{x}$\cite{I92}, but a scalar in the parameter $t$. This means that although
it makes sense to put a $\delta^3(\underline{x},\underline{x'})$ on the right hand
side of the Poisson bracket Eq.\ (\ref{GR3b})---where the insertion of the comma in
the notation $\delta^3(\underline{x},\underline{x}')$ indicates that it is a scalar in
$\underline{x}$ but a density\footnote{In standard canonical general relativity, in
the Poisson bracket $\{h_{ij}(\underline{x}),\pi^{kl}(\underline{x}^\prime)\}=
\delta_{(ij)}^{kl}\delta^3(\underline{x},\underline{x}')$, the quantity
$\pi^{kl}(\underline{x}^\prime)$ is a tensor {\em density} on $\Sigma$ of the
appropriate weight, whereas $h_{ij}(\underline x)$ is just a tensor field.}of weight 1
in $\underline{x'}$---it would not be correct to add the term $\delta(t,t')$ which is
a density in $t'$. Thus, if $u^{ij}(t,\underline{x})$ is a test density in both
variables $t$ and $\underline{x}$, and if $v_{kl}(t,\underline{x})$ is a density in
$t$ but a function in $\underline{x}$, then the smeared version of Eq.\ (\ref{GR3b})
is
\begin{equation}
\{h(u),\pi(v)\}=\int dt\,\alpha(t)\int d^3\underline{x}\, u^{ij}(t,\underline{x})
v_{ij}(t,\underline{x}) .
\end{equation}
We shall discuss next the physical meaning of the quantity $\alpha(t)$. We note
however that it can be regarded as a time-dependent analogue of the dimensioned
parameter $\tau$ which should appear on the right hand side of the {\em history\/}
version of Poisson brackets, as we have showed in \cite{Sav99}.

\paragraph{Some comments on $\alpha(t)$.}
We have been long discussing in work so far \cite{Sav99,SA00,Sav01}, the essential
difference between the {\em internal and external modes of time\/}, and in particular
the way they appear in the histories theory scheme. Ever since their original
construction\cite{Sav99}, the interpretation of the `two types of time' has served as
the key tool to further the particular histories theory (`History Projection
Operator') formalism, originally presented by Isham\cite{cji94} and Isham {\em et
al\/}\cite{HPOHis}.

In the past, we have attempted to present parts of the conceptual issues involved; yet
this will be the subject of a future work, that it will mainly involve presenting in a
detailed way these novel ideas about the concept of time. However, we cannot avoid
here some comments on these issues, as it is the first time that the mathematical
structure of the theory enables an immediate comparison between the `internal' and the
`external' pictures of the theory.

First, we can  think of the function $\alpha(t)$ as follows. In the case of a single
particle we can write the canonical symplectic form $\omega_t = dp_t \wedge dq_t$  on
the phase space $\Gamma_t$, for each moment of time $t$. Then the history symplectic
form for this system is defined by integrating
\begin{equation}
\Omega = \int d \mu(t) dp_t \wedge dq_t \,,
\end{equation}
where $\mu(\cdot)$ is an {\em arbitrary} measure on the real line $\mathR$. For the
case of continuous-time paths, and when time is defined along the whole real axis
$\mathR$, we can write $d\mu(t) = dt/\alpha(t)$, where  $\alpha(t) $ is a density,
that is defined to have dimensions of time, so that the history observables  have the
same dimension as the canonical ones. Hence $\alpha(t) $ is naturally associated with
the notion of `time-averaging'. In particular,  the freedom to choose an arbitrary
function $\alpha(t)$ reflects the freedom of the histories construction, to
arbitrarily select the `weight' by which each moment of time will contribute to the
time-averaging of physical quantities.

The above comment is, however, conceptually distinct from the notion of different
time-parameters arising from different foliations: the freedom in the choice of
$\alpha(t)$ is present even in simple non-relativistic systems.

In all systems we have studied so far \cite{Sav99,SA00,Sav01}, the choice of
$\alpha(t)$ was practically of no consequence, and we chose to set it equal to a
constant. However, in the context of general relativity, $\alpha(t)$ has an additional
significance: if the history observables are to treat time and space coordinates in
the same footing---the choice we followed through all our work so far and in accord to
the `two modes of time' interpretation---the introduction of a density $\alpha(t)$ is
unavoidable.

We believe that this is related to the interplay between canonical formalism and
covariant formalism, as they have appeared naturally intertwined in the histories
formalism once the introduction of two types of time transformation was made. Indeed,
the definition of the action operator $S$ in \cite{Sav99} already establishes an
interplay between Lagrangian formalism and Hamiltonian formalism, as $S$ is defined in
analogy to the classical Hamilton-Jacobi action functional. In future work we shall
argue that, the canonical and covariant descriptions implicitly involve a
correspondence to the `external' and `internal' time distinction, as it has been
presented so far in the histories theory. Hence, the Lagrangian and the Hamiltonian
formalism refer to {\em different\/} treatment of the two modes of time, even though
the two formalisms coincide at the level of the equations of motion.

In general relativity there exists already an implied distinction at the level of
symmetries, as the spacetime diffeomorphism group is different from the group of the
canonical constraints.

\bigskip
\bigskip

\subsection{The Dirac Algebra of Constraints}
The construction above leads naturally to a one-parameter family of Dirac
super-hamiltonians $t\mapsto{\cal{H}}_{\bot} ( t, \underline{x})$ and super-momenta
$t\mapsto {\cal{H}}_i ( t, \underline{x} )$. In the standard canonical approach to
general relativity\cite{ADM,Kuc91,I92}, the super-hamiltonian and super-momenta are
\begin{eqnarray}
{\cal{H}}_{\bot} &=& \kappa^2 h^{-1/2}(\pi^{ij}\pi_{ij}
- \frac{1}{2} (\pi_i{}^i)^2) - \kappa^{-2}h^{1/2} R          \\
{\cal{H}}^i  &=& - 2 {\nabla}_{\!\!j} \pi^{ij}, \label{ADMHi}
\end{eqnarray}
where $\kappa^2=8\pi G/c^2$ and $\nabla$ denotes the  spatial covariant derivative. We
note that both these quantities are spatial scalar densities, hence they can be
smeared with scalar quantities.

The history analogue of these expressions is
\begin{eqnarray}
{\cal H}_\perp(t,\underline x)&:=&\kappa^2
h^{-1/2}(t,\underline{x})(\pi^{ij}(t,\underline{x})\pi_{ij}(t,\underline{x})
- \frac{1}{2} (\pi_i{}^i)^2(t,\underline{x})) -\\
&&\kappa^{-2}h^{1/2}(t,\underline{x})
R(t,\underline{x}) \label{HperpHis}\\
\hspace{-1cm} {\cal{H}}^i(t,\underline x)&:=& - 2 {\nabla}_{\!\!j}
\pi^{ij}(t,\underline x).
\end{eqnarray}

For each choice of the weight function $\alpha$, these quantities on
$\mathR\times\Sigma$ satisfy the history version of the Dirac algebra
\begin{eqnarray}
\hspace{-1cm}\{ {\cal{H}}_i (t,\underline{x})\,, {\cal{H}}_j ( t^{\prime} ,
{\underline{x}} ^{\prime} )\}\ &=& - {\cal{H}}_j (t,\underline{x}) \,\delta ( t ,
t^{\prime})\alpha(t')\, {\partial^{x^{\prime}}}\!\!_{i} \,\delta^3 ( \underline{x}
,\underline{x}^{\prime}) \nonumber \\
 &&+ {\cal{H}}_i (t,\underline{x})\, \delta ( t , t^{\prime})\alpha(t')\,
{\partial^x}\!\!_{j}\, \delta^3 ( \underline{x} ,
\underline{x}^{\prime})  \label{Diracsm1}\\
\hspace{-1cm}\{ {\cal{H}}_i (t,\underline{x})\,, {\cal{H}}_{\bot} ( t^{\prime} ,
{\underline{x}}^{\prime} ) \} &=&  {\cal{H}}_{\bot} (t,\underline{x})\, \delta ( t ,
t^{\prime})\alpha(t')\, {\partial^{x^{\prime}}}\!\!_i \,\delta^3
( \underline{x} , \underline{x}^{\prime}) \label{Diracsm2} \\
\hspace{-1cm}\{ {\cal{H}}_{\bot} (t,\underline{x})\,, {\cal{H}}_{\bot} (
{\underline{x}}^{\prime} , t^{\prime} )\} &=&  h^{ij}\! (t,\underline{x})
\,{\cal{H}}_i (t,\underline{x})\,\delta ( t , t^{\prime})\alpha(t')\,
{\partial^{x^{\prime}}}\!\!_{j}\, \delta^3 ( \underline{x} ,
\underline{x}^{\prime}) \nonumber  \\
 &&\!- h^{ij}\! ( t^{\prime} , {\underline{x}} ^{\prime} )\, {\cal{H}}_i (
t^{\prime} , {\underline{x}} ^{\prime} )\,\delta ( t ,
t^{\prime})\alpha(t')\,{\partial^x}\!\!_{j}\, \delta^3 ( \underline{x} ,
\underline{x}^{\prime}).  \label{Diracsm3}
\end{eqnarray}
The smeared form of the super-hamiltonian ${\cal{H}}_{\bot} (t,\underline{x})$ and the
super-momentum ${\cal{H}}_i (t,\underline{x})$ history quantities are defined using as
their smearing functions a scalar function $L$, and a spatial vector field $L^i$ in
the following way
\begin{eqnarray}
{\cal{H}} (L) := \int d^3 \underline{x} \int dt\,\alpha(t)^{-1} L(t,\underline{x})
{\cal{H}}_{\bot} (t,\underline{x})     \\
{\cal{H}} (\vec{L}) := \int d^3 \underline{x} \int dt\,\alpha(t)^{-1} L^i
(t,\underline{x}){\cal{H}}_i (t,\underline{x}).
\end{eqnarray}
Hence we can write the smeared form of the `Hamiltonian' as
\begin{eqnarray}
H(L, \vec{L} )&=& \int d^3 \underline{x} \int dt\,\alpha(t)^{-1}\, \Big{(} L
(t,\underline{x}) {\cal{H}}_{\bot} ( t , \underline{x} ) + L^i
(t,\underline{x}){\cal{H}}_i (t,\underline{x}) \Big{)}
\\ \nonumber
 &=& {\cal{H}} (L)+ {\cal{H}} (\vec{L}).
\end{eqnarray}

The smeared form of this history version of the Dirac algebra is
\begin{eqnarray}
\!\!\{ {\cal{H}}[ \vec{L} ]\,, {\cal{H}}[\vec{L^{\prime}}]\}\
&=& {\cal{H}} [ \vec{L} \,\,, \vec{L^{\prime}} ] \label{NiN'i} \\
\!\!\{ {\cal{H}} [ \vec{L} ] \,, {\cal{H}} [ L ] \} &=& {\cal{H}} [
{L_{\vec{L}}} L ]  \label{H(NiN)} \\
\!\!\{ {\cal{H}}[ L ]\,, {\cal{H}} [L^{\prime}] \} &=& {\cal{H}} [ \vec{K}],
\label{H(NN')}
\end{eqnarray}
where in Eq.\ (\ref{H(NN')}) we have $K^i := h^{ij}(L \partial_j L' - L' \partial_j
L)$, with $i=1,2,3$.

\bigskip

\subsection{The representation of the group ${\rm Diff}(M)$}
We note that this smeared form Eqs.\ (\ref{NiN'i}--\ref{H(NN')}) of the Dirac algebra
is the analogue of the {\em internal\/} Poincar\'{e} group of the histories quantum
field theory, in the sense that it does not affect the external time label $t$. We
shall now see that there is also an analogue of the external Poincar\'e group---namely
the group of spacetime diffeomorphisms.

Since we have considered a one-parameter family of Riemannian metrics $h_{ij}$ on the
three-surface $\Sigma$, we can now identify $\mathR\times\Sigma$ as the spacetime $M$.
The critical observation here is that we can write a representation of the spacetime
diffeomorphisms group ${\rm Diff}(M)$ on a suitable extension of the canonical history
space $\Pi$, which will also carry the representation of the history version of the
Dirac algebra discussed above.

In order to demonstrate this statement we start by postulating the `covariant' Poisson
brackets, on the extended history space $\Pi^{ext}$
\begin{eqnarray}
\{g_{\mu\nu}(X)\,, \,g_{\alpha\beta}(X^{\prime})\}&=& 0 \label{covgg}  \\
\{\pi^{\mu\nu}(X)\,, \,\pi^{\alpha\beta}(X^{\prime})\} &=& 0\label{covpipi} \\
\{g_{\mu\nu}(X)\,, \,\pi^{\alpha\beta}(X^{\prime})\} &=& \delta_{(\mu\nu
)}^{\alpha\beta} \,\delta^4 (X, X^{\prime}) , \label{covgpi}
\end{eqnarray}
where $X$ is a point in the spacetime $M$, and where $g_{\mu\nu}(X)$ is a four-metric
that belongs to the space of Lorentzian metrics $L {\rm Riem}(M)$, and
$\pi^{\mu\nu}(X)$ is the conjugate variable. We have defined
${{\delta}_{(\mu\nu)}}^{\alpha\beta} := \frac{1}{2}({\delta_\mu}^\alpha
{\delta_\nu}^\beta + {\delta_\mu}^\beta {\delta_\nu}^\alpha )$.

In previous applications of the histories formalism we have defined the `Liouville'
function $V$ as the generator of time translations with respect to the `external' time
$t$ that appears as a kinematical ordering parameter that distinguishes between past,
present and future \cite{Sav99, SA00, Sav01}. In the present case, in analogy with
previous history constructions, we can define the `Liouville' function $V_W$
associated with any vector field $W$ on $M$ as
\begin{equation}
    V_W:=\int \!d^4X \,\pi^{\mu\nu}(X)\,{\cal L}_W g_{\mu\nu}(X) \label{Vw}
\end{equation}
where ${\cal L}_W$ denotes the Lie derivative with respect to $W$. This is the direct
analogue of the expression that is used in the normal canonical theory for the
representations of spatial diffeomorphisms.

The fundamental result is that these generalised Liouville functions $V_W$, defined
for any vector field $W$ as in Eq.\ (\ref{Vw}),  satisfy the Lie algebra of the {\em
spacetime diffeomorphisms group\/} ${\rm Diff}(M)$
\begin{equation}
\{\, V_{W_1}\,, V_{W_2}\,\} = V_{ [ W_1 , W_2 ]},
\end{equation}
where $[ W_1 , W_2 ]$ is the Lie bracket between vector fields $W_1$ and $W_2$ on the
manifold $M$.

Now, the aim is to show how the use of the covariant brackets Eqs.\
(\ref{covgg}--\ref{covgpi}) leads to an augmented history space $\Pi^{ext}$, in which
we can recover the history Dirac algebra constructed in the previous section.

The first step in recovering the history algebra Eqs.\ (\ref{GR1b}--\ref{GR3b}) is to
choose a foliation ${\cal F}: \mathR\times\Sigma \longrightarrow M$. Then we define
the spatial parts of the pull-back of $g_{\mu\nu}(X)$ to $\mathR\times\Sigma$ by
${\cal F}$ as
\begin{equation}
h_{ij}(t,\underline{x}) := {\cal{F}}^{\mu}_{,i}(t,\underline{x})\,
{\cal{F}}^{\nu}_{,j}(t,\underline{x})\, g_{\mu\nu}({\cal{F}}(t,\underline{x}))
\label{Pullbackh}
\end{equation}
where ${\cal{F}}^{\mu}_{,i}(t,\underline{x}):={\partial}_i
({\cal{F}}^{\mu}(t,\underline{x}))$.

For a fixed $g$, we can choose the foliation to be spacelike\footnote{For an
appropriate topology on $L{\rm Riem}(M)$, this spacelike character will be maintained
for some open neighborhood of the Lorentzian metric $g$. However, this foliation will
fail to be spacelike for certain other Lorentzian metrics on $M$. This feature is not
important at the level of the classical theory we are discussing here; however it can
be expected to be a non-trivial issue in the quantum theory.} in the sense that
$t\mapsto h_{ij}(t,\underline{x})$ is a path in the space of Riemannian metrics on
$\Sigma$.

Next, we need to pull-back the conjugate variable $\pi^{\alpha\beta}(X)$ to
$\mathR\times\Sigma$ also. For this purpose, we lower the indices and define the field
$\pi_{\alpha\beta}(X)= g_{\gamma\alpha}(X)\,g_{\zeta\beta}(X)\, \pi^{\gamma\zeta}(X)$.
Hence, using the Poisson brackets Eqs.\ (\ref{covgg}--\ref{covgpi}), we get the
relations
\begin{eqnarray}
\{g_{\mu\nu}(X)\,, \,g_{\alpha\beta}(X^{\prime})\}&=& 0 \label{covgg}  \\
\{\pi_{\mu\nu}(X)\,, \,\pi_{\alpha\beta}(X^{\prime})\} &=& 0\label{covpipi} \\
\{g_{\mu\nu}(X)\,, \,\pi_{\alpha\beta}(X^{\prime})\} &=&
{g_{(\mu\alpha}}\,{g_{\nu)\beta}(X)} \;\delta^4 \!(X,\, X^{\prime}) \label{covgpi}
\end{eqnarray}
where $g_{(\mu\alpha}g_{\nu)\beta}(X):= \frac{1}{2} (g_{\mu\alpha}(X)\,g_{\nu\beta}(X)
+ g_{\nu\alpha}(X)\,g_{\mu\beta}(X))$. We must now pull back these equations Eq.\
(\ref{covgg}--\ref{covgpi}) to $\mathR\times\Sigma$ using the foliation ${\cal
F}:\mathR\times\Sigma\rightarrow M$.

Here, it is important to notice that, since $\delta^4(X,X')$ is a {\em density\/} in
the variable $X'$, we have---in coordinates $(t,\underline{x})$ adapted to the split
$\mathR\times\Sigma$---the relation
\begin{equation}
    \delta^4({\cal F}(t,\underline{x}),{\cal F}(t',\underline{x}'))=
        K(t',\underline{x}') \delta(t,t')\delta^3(\underline{x},\underline{x}') \,,
\end{equation}
where $K(t',\underline{x}')$ is an appropriate power of the Jacobian of the
diffeomorphism ${\cal F}:\mathR\times\Sigma\rightarrow M$. However, since
$\pi_{\alpha\beta}(X')$ is a tensor density on $M$ of the same weight as the second
variable in $\delta^4(X,X')$, we consider the quantity $\pi_{ij}(t,\underline{x})$
defined by
\begin{equation}
\pi_{ij}(t,\underline{x}) := K(t,\underline{x})
{\cal{F}}^{\mu}_{,i}(t,\underline{x})\, {\cal{F}}^{\nu}_{,j}(t,\underline{x})\,
\pi_{\mu\nu}({\cal{F}}(t,\underline{x})).
\end{equation}

These new quantities $h_{ij}(t,\underline{x})$, and $\pi_{ij}(t,\underline{x})$
satisfy the Poisson brackets
\begin{eqnarray}
\{ h_{ij}(t,\underline{x}) \,, h_{kl}(t^{\prime},\underline{x^{\prime}}) \}&=& 0
\label{canhh}\\
\{ \pi_{ij}(t,\underline{x}) \,, \pi_{kl}(t^{\prime},\underline{x^{\prime}}) \}&=& 0
\label{canpp}\\
\{ h_{ij}(t,\underline{x}) \,, \pi_{kl}(t^{\prime},\underline{x^{\prime}}) \}&=&
{h_{(ik}}(t,\underline{x})\,{h_{j)l}(t,\underline{x})} \, \delta(t,
t^{\prime})\,\delta^{3}(\underline{x} \,, \underline{x^{\prime}}) \label{canhp}
\end{eqnarray}
where we have defined ${h_{(ik}}\,{h_{j)l}}:= \frac{1}{2}
(h_{ik}\,h_{jl} + h_{jk}h_{il} )$.

Finally, we define
\begin{equation}
 \pi^{kl}(t,\underline{x}):= \alpha(t)\;h^{ka}(t,\underline{x})\,h^{lb}(t,\underline{x})\,
  \pi_{ab}(t,\underline{x}),
\end{equation}
where $h^{ka}(t,\underline{x})$ is the inverse of the Riemannian metric
$h_{ka}(t,\underline{x})$ on $\Sigma$, for each $t$. Hence we have regained the
canonical Poisson brackets Eqs.\ (\ref{GR1b}--\ref{GR3b}) for the histories canonical
treatment in section $3.1$.

\subsection{The extended state space $\Pi^{ext}$ }
In the previous section, although we referred to an augmented version $\Pi^{ext}$ of
the history space $\Pi$, in which both a representation of the group of spacetime
diffeomorphisms and of the Dirac algebra of constraints exist, there was no need, for
the purposes of that section, for a detailed presentation of the extended state space.
However, now we shall present the augmented canonical history algebra, to show that
the physical predictions of the previous sections hold, and to further examine
possible interesting implications of the new  construction for the histories version
of general relativity.

When we go from the covariant Poisson brackets, that involve the spacetime metric
$g_{\mu \nu} (X)$, to the Poisson brackets that involve the paths of Riemannian metric
$h_{ij}(t, \underline{x})$ , we ignore the quantities that in the standard $3+1$
decomposition correspond to the lapse function and the shift vector. When we do take
them into account it amounts into a space of paths on an {\em extended state space}
$\Pi^{ext}$.

To this end, we first recall that the choice of a foliation ${\cal F}$ enables us to
decompose the spacetime metric $g_{\mu\nu}$ in a coordinate system adapted to the
foliation ${\cal F}$, as for instance
\begin{eqnarray}
g_{00}(t,\underline{x}) &=& g_{\mu \nu}({\cal F}(t,\underline{x})) \,\dot{{\cal
F}}^{\mu}(t,\underline{x}) \,\dot{{\cal
F}}^{\nu}(t,\underline{x}) \\
g_{0i}(t,\underline{x}) &=& g_{\mu \nu}({\cal F}(t,\underline{x})) \;\dot{{\cal
F}}^{\mu}(t,\underline{x}) \;{\cal F}_{,i}^{\nu}(t, \underline{x}),
\end{eqnarray}
where we write $\dot{{\cal F}}(t, \underline{x}) = \partial_t {\cal F}(t,
\underline{x})$.

If the unit, timelike vector field $n^{\mu}$ is normal to the foliation, we can write
\begin{equation}
\dot{{\cal F}}^{\mu}(t,\underline{x}) = \tilde{N}(t,\underline{x}) \,n^{\mu}({\cal
F}(t,\underline{x})) + {\cal F}^{\mu}_{,i}(t,\underline{x}) \,\tilde{N}^i(t,
\underline{x}).
\end{equation}
The above equation defines the lapse function $\tilde{N}$ and the shift vector
$\tilde{N}^i$, associated with the foliation ${\cal F}$ and the metric $g_{\mu
\nu}(X)$. It is a standard result that the {\em inverse metric} $g^{\mu\nu}$ can be
written in the coordinate system adapted to the foliation as
\begin{eqnarray}
g^{00}(t,\underline{x})&=&   \tilde{N}^2(t,\underline{x}) \\
g^{0i}(t,\underline{x}) &=&   \tilde{N}^i(t,\underline{x}) \\
g^{ij}(t,\underline{x}) &=&  h^{ij}(t,\underline{x})\,,
\end{eqnarray}
where $h^{ij}$ is the inverse of $h_{ij}$, as in Eq.\ (\ref{Pullbackh}).

Next we write the Poisson brackets in terms of the inverse metric $g^{\mu \nu}$ and
the field $\pi_{\rho \sigma}$ as
\begin{equation}
\{g^{\mu \nu}(X), \pi_{\rho \sigma}(X') \} = - \delta^{\mu \nu}_{(\rho \sigma)}
\delta^4(X,X') \,.
\end{equation}
From this expression it is easy to check that the quantities $p(t,\underline{x})$ and
$p_i(t,\underline{x})$ defined as
 \begin{eqnarray}
  p(t,\underline{x}) &:=& - 2  \tilde{N}(t,\underline{x})\,
K(t,\underline{x})\,\pi_{\mu \nu}({\cal F}(t,\underline{x}))\, \dot{{\cal
F}}^{\mu}(t,\underline{x}) \,
\dot{{\cal F}}^{\nu}(t,\underline{x}) \\
 p_i(t,\underline{x}) &:=& -K(t,\underline{x}) \, \pi_{\mu \nu}({\cal
 F}(t,\underline{x}))\,
\dot{{\cal F}}^{\mu}(t,\underline{x})\, {\cal F}_{,i}^{\nu}(t,\underline{x})
\end{eqnarray}
are conjugate momenta to the history objects $N(t,\underline{x}) : = \alpha(t)
\tilde{N}(t, \underline{x})$ and $N^i(t,\underline{x}) :=\alpha(t)
\tilde{N}^i(t,\underline{x}) $ \footnote{The objects $\tilde{N}(t,\underline{x})$ and
$\tilde{N}^i(t,\underline{x})$ are densities with respect to reparameterisations of
the $t$ label, hence the association $t\mapsto \tilde{N}(t,\underline{x})$ does {\em
not\/} correspond to a path in the space of scalar fields on $\Sigma$. This is the
reason we prefer to use as history canonical variables the objects
$N(t,\underline{x})$ and $N^i(t,\underline{x})$, that do correspond to a path on the
space of scalar fields or vector fields on $\Sigma$ respectively.} respectively, in
the sense that they satisfy the Poisson brackets equations
\begin{eqnarray}
\{N(t,\underline{x}), p(t',\underline{x}')\} &=& \alpha(t) \delta(t,t')
\delta^3(\underline{x}', \underline{x'}) \label{cansNp}\\
\{N(t,\underline{x}), N(t', \underline{x'}) \} &=& 0 \label{cansNN}\\
\{p(t,\underline{x}), p(t', \underline{x'}) \} &=& 0 \label{canspp} \\
 \{ N^i(t, \underline{x}), p_j(t',\underline{x}') \} &=& \delta^i_j\alpha(t)
\delta(t,t')\delta^3(\underline{x}', \underline{x'}) \label{canvNp} \\
\{N^i(t,\underline{x}), N^j(t', \underline{x'}) \} &=& 0 \label{canvNN} \\
\{p_i(t,\underline{x}), p_j(t', \underline{x'}) \} &=& 0 \,, \label{canvpp}
\end{eqnarray}
and that all quantities $N,N^i,p$ and $p_i$ have vanishing Poisson brackets with
$p_{ij}$ and $h^{ij}$.

Hence, given a foliation ${\cal F}$, one can write the covariant Poisson brackets in
terms of objects that represent paths into an {\em extended phase space} having as
basic Poisson brackets Eqs.\ (\ref{canhh}--\ref{canhp}) and Eqs.\ (
\ref{cansNp}--\ref{canvpp}).

It is important to emphasise here that, because the generators ${\cal
H}_\perp(t,\underline x)$ and ${\cal{H}}^i(t,\underline x)$ of the history Dirac
algebra Eqs.\ (\ref{Diracsm1}--\ref{Diracsm3}), trivially commute with the additional
variables of the extended history algebra, we recover exactly the history version of
the Dirac algebra, as it was originally defined in section $3.1$. Therefore, on the
extended history space $\Pi^{ext}$ we have a representation of the Dirac algebra {\em
together\/ with\/} a representation of the spacetime diffeomorphisms group ${\rm
Diff}(M)$.

\subsection{The history space $\Pi$ is the $3+1$-decomposition state space }
In the standard Hamiltonian analysis of the Einstein-Hilbert action, one goes from the
extended phase space (involving the lapse function $N$, shift vector $N^i$, and their
conjugate momenta $p$ and $p_i $), to the state space (containing only three-metrics
$h_{ij}$ and their conjugate momenta $\pi^{kl}$), by imposing as first-class
constraints the vanishing of $p(t,\underline{x})$ and $p_i (t,\underline{x})$.

In histories theory, and in analogy to the parameterised systems algorithm we
established in \cite{SA00}, we implement the history analogues of the canonical
constraints in the extended history phase space $\Pi^{ext}$, i.e., we impose the
conditions
\begin{eqnarray}
p(t,\underline{x}) &=& 0  \label{conp}\\
p_i(t,\underline{x}) &=& 0. \label{conpi}
\end{eqnarray}
Together with the vanishing of the super-hamiltonian ${\cal{H}}_{\bot} ( t,
\underline{x})$ and the super-momentum ${\cal{H}}_i ( t, \underline{x} )$, the above
equations form a set of first-class constraints.

We impose the constraints Eqs.\ (\ref{conp}--\ref{conpi}) to vanish on the constraint
surface $C_h$, which is essentially the space of all paths $t\mapsto C$, from the real
line $\mathR$ to the constraint surface $C$ of standard canonical theory\cite{SA00}.

We then consider the space of orbits $C_h/\sim$  with respect to the action of the
symplectic transformations generated by the constraints \footnote{We reserve the name
of ``reduced phase space'' for the space obtained by the implementation of all
constraints.}.

The constraint functions $p$ and $p_i$ commute with $h_{ij}$ and $\pi^{ij}$, hence the
symplectic transformations generated by these constraints leaves $h_{ij}$ and
$\pi^{ij}$ invariant. Hence, we recover the history space $\Pi$---originally defined
in section $3.1$---as being the space of orbits \footnote{In most discussions of the
Hamiltonian description of general relativity, the constraint equations $p = 0$ and
$p_i = 0$ are imposed early in the discussion hence the `reduced state space'---which
is defined as the space of orbits of the constraints' action on the constraint
surface---consists of the three-metrics $h_{ij}$ and their conjugate momenta
$\pi^{ij}$. and it is referred to as the state space of general relativity.}
corresponding to these constraints.

Furthermore, it is interesting to see how the generators $V_W$ of the diffeomorphisms
group can be projected into $\Pi$. To this end, we first write the above set of
constraints with the equivalent covariant expression
\begin{equation}
\Phi(k) = \int d^4X \pi_{\mu \nu}(X) t^{\mu}(X) k^{\nu}(X)\,,
\end{equation}
where $t^{\mu}(X): = \dot{{\cal F}^{\mu}}({\cal F}^{-1}(X))$ is the vector field
corresponding to the parameter $t$ of the foliation, and $k^{\mu}(X)$ is an arbitrary
vector field that serves as a smearing function. We then impose the constraint
$\Phi(k) = 0$ on the constraint surface.

Then, the commutator of the generators $V_W$ with the constraint $\Phi(k)$ is
\begin{eqnarray}
\hspace{-0.2cm}\{V_W, \Phi(k) \} \hspace{-0.3cm} &=& \hspace{-0.3cm} \int\! d^4X
({\cal L}_W\pi)_{\mu \nu} k^{\mu} t^{\nu}\hspace{-0.1cm} = \hspace{-0.1cm}\int\! d^4X
[{\cal L}_W(\pi_{\mu \nu} k^{\mu} t^{\nu})\! - \!\pi_{\mu \nu} {\cal
L}_W(k^{\mu} t^{\nu})]  \\
\hspace{-0.2cm}&=&\hspace{-0.2cm}- \int\! d^4X [\frac{1}{\sqrt{-g}}({\cal L}_W
\sqrt{-g}) \pi_{\mu \nu}k^{\mu} t^{\nu} - \frac{1}{\sqrt{-g}}
\partial_{\rho}(\sqrt{-g} W^{\rho}) \pi_{\mu \nu} k^{\mu} t^{\nu} \nonumber \\
  &&\hspace{-0.15cm}- \,\pi_{\mu \nu}({\cal L}_W k)^{\mu} t^{\nu} + k^{\mu}
({\cal L}_W t)^{\nu})]\,. \label{Phi(k)}
\end{eqnarray}
Note here that $\pi_{\mu \nu}$ is a tensor density. The first term of the expression
Eq.\ (\ref{Phi(k)}) is equal to $\Phi(\frac{1}{\sqrt{-g}}({\cal L} \sqrt{-g})k)$, the
second to $ \Phi(\frac{1}{\sqrt{-g}}
\partial_{\rho} (\sqrt{-g} W^{\rho}) k)$ and the third  to $\Phi({\cal L}_Wk)$, and all
three terms vanish on the constraint surface. However, the fourth term vanishes if and
only if
\begin{equation}
{\cal L}_Wt^{\mu}(X) := [W,t]^{\mu}(X) = f(X) t^{\mu}(X) \,,
\end{equation}
for some scalar function $f$. The above expression implies that the vector field $W$
preserves the foliation in the sense that its Lie bracket with the transverse field
$t^{\mu}$ yields a field in the same direction. Written in a coordinate system adapted
to the foliation this condition implies that the component $W^0(t,x)$ is a function of
$t$ only.

Therefore we conclude that, the histories space $\Pi$ carries a representation of the
sub group of foliation-preserving diffeomorphisms $ G_{\cal F}$, of the
diffeomorphisms group ${\rm Diff}(M)$. The generators of this sub-group can be written
as
\begin{equation}
V_W = \int \!d^3 \underline{x} \int\! dt \;\pi^{ij}(t,\underline{x})
[W^0(t)\dot{h}_{ij} + {\cal L}_{\overrightarrow{W}} h_{ij}](t,\underline{x}),
\end{equation}
where $\overrightarrow{W}$ denotes vector fields that are horizontal to the foliation.

\bigskip

\paragraph{Histories equations of motion.}
For the vector field $(W^0(t)\! = \!1$ , $\overrightarrow{W}\! =\! 0)$, we write the
Liouville function $V$ as
\begin{equation}
V:=\int d^3 \underline{x} \int dt\, \pi^{ij}(t,\underline{x}){\partial\over\partial
t}h_{ij}(t,\underline{x}).
\end{equation}
Following the histories methods of \cite{Sav99} we further define the histories action
functional $S$ as
\begin{eqnarray}
S &=& \int d^3 \underline{x} \int dt \left \{
\pi^{ij} (t,\underline{x})\, {\dot{h}}_{ij}(t,\underline{x}) - {\cal{H}} (\alpha L) -
{\cal{H}} (\alpha\vec{L}) \right\}  \\
&=& V - H ( \alpha L \,,\alpha\vec{L} ) \,,\label{Shis}
\end{eqnarray}
where $L$ and $\vec{L}$ are appropriate smearing functions. We notice here that the
quantities $L$ and $\vec{L}$ are not the original lapse function and shift vector,
since all trace of them was lost when passing from the extended state space
$\Pi^{ext}$ to $\Pi$.

It is easy to show that the usual dynamical equations for the canonical fields
$h_{ij}$ and $\pi^{ij}$ are equivalent to the history Poisson bracket equations
\begin{eqnarray}
\{ S \,, h_{ij} (t,\underline{x}) \}\, (\gamma_{cl}) &=& 0   \\
\{ S \,, \pi^{ij} (t,\underline{x}) \}\, (\gamma_{cl}) &=& 0
\end{eqnarray}
where $S$ is defined in Eq.\ (\ref{Shis}). The path $\gamma_{cl}$ is a solution of the
classical equations of motion, and therefore corresponds to a spacetime metric that is
a solution of the Einstein equations.

\subsubsection{ State space reduction.}
Next we should employ the algorithm we used in \cite{SA00} to treat parameterised
systems. Again  the `history constraint' surface $C_h$ is the space of paths from
$\mathR$ to the canonical constraint surface $C$, which is defined from the
requirement that the constraints should vanish for all times $t$.

We then study the action of the constraints by symplectic transformations on $C_h$.
The reduced space of histories $\Pi_{red}$ is the space of the orbits that are
obtained by the action of the constraints ({\em i.e.}, equivalence classes of points
of $C_h$ that are related by a constraint transformation):
\begin{equation}
\Pi_{red} = C_h / \sim
\end{equation}
In fact  $\Pi_{red}$ is isomorphic to the space of continuous paths on $\Gamma_{red}$.

In order for a function on $ \Pi $ to be a physical observable ({\em i.e.,\/} it can
be projected into a function on $ {\Pi}_{red}$), if and only if it commutes with the
constraints on the constraint surface. We shall now discuss the extent to which the
generators $V_W$ of the restricted spacetime  diffeomorphisms group $G_{\cal{F}}$,
that is represented in $\Pi$ satisfy this condition. Indeed,
\begin{eqnarray}
\{V_W, {\cal H}(\alpha L) \} &=& {\cal H}(\alpha {\cal L}_W L)  \label{Vw,HL} \\
\{ V_W, {\cal H}(\alpha \overrightarrow{L}) \} &=& V_{[W,\overrightarrow{L}]}
\hspace{0.5cm}= {\cal H}(\alpha [W, \overrightarrow{L}]) \,. \label{Vw,HLi}
\end{eqnarray}
The generators of the restricted spacetime diffeomorphisms $V_W$ clearly commute with the
super-hamiltonian ${\cal{H}}$ on the constraint surface $C_h$.

However, $V_W$ only commutes with ${\cal{H}}( {\vec{N}})$ on $C_h$ if $[W,\vec{N}]$ is
a spatial vector field, {\em i.e.}, if the diffeomorphisms generated by $W$ preserve
the spatial nature of $\vec{N}$. This is equivalent to the condition that the
diffeomorphisms generated by $W$ preserve the foliation.

Hence, amongst all spacetime diffeomorphisms, it is only the Liouville function
$V$---the time translations generator---that can be non-trivially
projected on the reduced phase
space (horizontal diffeomorphisms vanish on $\Pi_{red}$)
\begin{equation}
V_{\lambda} = \int\! dt \,\lambda(t) \int \!d^3\! \underline{x} \,\pi^{ij}
\frac{\partial} {\partial t} h_{ij}.
\end{equation}
for any function $\lambda$.

We note that the function $V_{\lambda}$ generates time reparametrisations of the
parameter $t$ on $\Pi_{red}$ \cite{SA00}. Hence, of all spacetime diffeomorphisms, it
is only the generators of time reparametrisations that is defined on the reduced phase
space $\Pi_{red}$.

\section{Conclusions}
We have showed how the recent development in introducing the distinction between time
as a causal ordering parameter, and as an evolution parameter in dynamics
\cite{Sav99}, leads to the construction of a history version of general relativity in
which there emerges a new relation between the group structures associated with the
normal Lagrangian and Hamiltonian approaches. In particular, we have showed that in
this histories version of canonical general relativity there exists a representation
of the spacetime diffeomorphisms group ${\rm Diff}(M)$, together with a history
analogue of the Dirac algebra of constraints.

However, various important issues arise. The immediate one to be addressed is that the
history algebra Eqs.\ (\ref{GR1b}--\ref{GR3b}) depends on the choice of a Lorentzian
foliation. This leads to two distinct questions. First, what is the degree to which
physical results depend upon this choice? The solutions to the equations of motion for
each choice allow us to construct different 4-metrics. If different descriptions are
to be equivalent, two distinct 4-metrics should be related by a spacetime
diffeomorphism. We should therefore establish that the action of the spacetime
diffeomorphisms group  intertwines between constructions corresponding to different
choices of the foliation. This involves considering state space histories
corresponding to arbitrary choices of foliation.

Second, and perhaps more important, is to question the notion of a spacelike foliation
itself. Since the spacetime causal structure is a dynamical object, the notion of a
foliation being {\em spacelike\/} has meaning only {\em after\/} the solution to the
classical equations of motion has been selected. However, in the histories description
we do not just use a single solution of the classical equations of motion (indeed,
many of the possible histories are not solutions at all), and in these circumstances
the notion of a `spacelike' foliation loses its meaning.

The issues mentioned above are fundamental in the treatment of general relativity.
Once they have been resolved, further applications will be technically
straightforward: for instance, the appearance of Noether's theorem in the histories
formalism. Of particular significance is the fact that the histories description
contains a mixture of Lagrangian and Hamiltonian structures. An interesting
application would be to apply the history ideas to the description of gravity in terms
of the Ashtekar variables.

The work presented here is only the beginning of a programme for constructing a
history theory of general relativity. What is necessary next, is to develop the
formalism to find a description that focuses on a manifestly covariant treatment of
the theory.

\vspace{2 cm}

\noindent{\large\bf Acknowledgements}

\noindent I would like to thank Charis Anastopoulos for a very fruitful interaction
and Karel Kuchar for useful discussions. I would like to especially thank Chris Isham
for his help on differential-geometric construction issues. I gratefully acknowledge
support from the L.D. Rope Third Charitable Settlement and from EPSRC GR/R36572 grant.

\end{document}